\documentclass[aps,11pt,prc,preprint,superscriptaddress,nofootinbib]{revtex4}
\usepackage[usenames]{color}
\usepackage{graphicx}
\usepackage{amsmath}
\usepackage{amsfonts}
\usepackage{amssymb}
\usepackage{mathrsfs}
\usepackage{bm}
\usepackage{verbatim}

\newcommand{\beq}{\begin{equation}}
\newcommand{\eeq}{\end{equation}}
\newcommand{\bea}{\begin{eqnarray}}
\newcommand{\eea}{\end{eqnarray}}

\newcommand{\chn}[3]{{{}^{#1}{#2}_{#3}}}
\newcommand{\cs}[2]{\chn{#1}{S}{#2}}
\newcommand{\cp}[2]{\chn{#1}{P}{#2}}

\graphicspath{{../}}

\begin{document}

\title{Chiral potential renormalized in harmonic-oscillator space}
\author{C.-J.~Yang}
\affiliation{Institut de Physique Nucleaire, IN2P3-CNRS, Universite Paris-Sud, F-91406
Orsay Cedex, France}
\email{yangjerry@ipno.in2p3.fr}
\date{\today }

\begin{abstract}
We renormalize the chiral effective field theory (EFT) potential in
harmonic-oscillator (HO) model space. The low energy constants (LECs) are
utilized to absorb not just the ultra-violet part of the physics due to the
cutoff, but also the infrared part due to the truncation of model space. We
use the inverse J-matrix method to reproduce the nucleon-nucleon (NN)
scattering phase shifts in the given model space. We demonstrate that by
including the NLO correction, the nucleon-nucleon scattering in the
continuum could be well reproduced in the truncated HO trap space up to
laboratory energy $T_{lab}=100$ MeV with number of HO basis $n_{max}$ as
small as 10. A perturbative power counting starts at subleading order is
adopted in this work, and how to extract the perturbative contribution is
demonstrated. Our work serves as the input to perform ab-initio calculations.
\end{abstract}

\pacs{12.39.Fe, 25.30.Bf, 21.45.-v, 21.60.Cs }
\maketitle

\vspace{10mm}

\section{Introduction}

\bigskip With the development of computational power and technique, the
input of ab-initio calculation---NN interaction---has became the one of the
main uncertainties in the few- and many-body calculations. Due to the fact
that calculations needed to be performed within a model space, an effective
and model-independent NN potential which converges fast enough within the
model space is required. Except for Fadeev-like approaches and, e.g., the
Monte-Carlo techniques and in general ab initio nuclear reaction approaches
that use wave functions with proper boundary conditions---which only
involves the ultraviolet cutoff---the model space is usually truncated in
both the ultraviolet ($\Lambda $) and infrared ($\lambda $) scales.
Therefore, before trusting the results one has to check carefully the
convergent pattern with respect to $\Lambda $ and $\lambda $.

In the past two decades, interaction deduced from effective filed theory
(EFT)\cite{We90,We91,Ordonez:1993-1995, Epelbaum, idaho, Epelbaummore,
reviews,reviews2,reviews3,NTvK05} has been developed and significant efforts
has been spent on the goal of providing good and model-independent
description of data. The bare NN interaction obtained in the continuum
cannot be directly applied into few- or many-body calculation since: (a) The
model space to perform calculations usually contains both ultraviolet and
infrared cutoffs. (b) Ultraviolet cutoff of the bare potential is too hard%
\footnote{%
The bare potential usually has an ultraviolet cutoff which is too high for
the results to converge within the limited number of basis.}. One common
treatment is to perform an unitary transformation (such as Lee-Suzuki or
similarity renormalization group (SRG)) to generate effective interactions
from the bare one\cite{uni}. However, in addition to loosing some level of
resolution---which is unavoidable when model space is reduced---the unitary
transformation also generates at least one additional scale, such as the
SRG\ flow parameter and the induced higher body force. Since the EFT power
counting is usually organized in the continuum with respect to $\Lambda $
alone, those extra cutoffs (due to $\lambda $ and other additional scales
from unitary transformation) could in principle destroy the power counting
after the transformation. Without a complete check, the interaction serves
as input will lose its model independent feature.

It is therefore desirable to build the effective interactions within a
limited number of basis in an alternative way. The philosophy of present
work is that, instead of renormalizing the interaction in the continuum
first and transforming it into a given model space later, hoping to find a
method of truncation (along with certain conditions) which does not affect
the model-independent feature of the original interaction, one performs the
renormalization directly in a given model space by utilizing the low energy
constants (LECs) in the EFT.

This direction has been advocated by the Arizona group \cite%
{Ia05,Ia07a,Ia07b,Ia10,Ji10,Ia10a,RJ12}, where the pionless potential is
direct renormalized in a given HO space with or without a physical trap. The
use of the physical trap allows one to connect the phase shifts ($\delta$)
to eigenvalues of matrix element by Busch formula\cite{busch, Luu}. Similar
applications to bosonic system is carried out in Ref.\cite{To}. An
alternative approach is to adopt the J-matrix formalism to relate $\delta$
to the eigenvalues\cite{oak}. Finally, an approach which deduces the
potential through Bloch-Horowitz equation---the HO-based effective theory
(HOBET)---is also explored\cite{hobet,hobet2}.

In this work the harmonic oscillator (HO) basis is adopted. We construct the
chiral EFT interaction directly in a given model space without the HO trap.
Similar to Ref.\cite{oak}, the renormalization of the EFT interaction under
the given infrared cutoff $\lambda $ and ultraviolet cutoff $\Lambda $ is
done through the inverse scattering J-matrix method\cite{jisp,jisp2,jisp3},
which enables a direct connection between the eigenstates in truncated HO
space to the NN scattering phase shifts. The truncated model space is
characterized by an ultraviolet cutoff $\Lambda \sim \sqrt{%
M(N_{_{max}}+3/2)\hbar \omega }$ and an infrared cutoff $\lambda \sim \sqrt{%
\frac{M\omega }{4(N_{max}+7/2)}}$. Here $M$ is the nucleon mass, $%
N_{_{max}}=2n_{\max }+l$, with $n_{\max }$ the maximum number of shells
included in the calculation and $l$ the angular momentum quantum number, $%
\omega $ is the oscillator frequency associated with the HO basis used. Note
that under the condition that the ultraviolet cutoff is saturated, detail
studies in Refs.\cite{infr,infr2,infr3} suggest that $\lambda =\sqrt{\frac{%
M\omega }{4(N_{max}+7/2)}}$ should be adopted, and $\Lambda =\sqrt{%
M(N_{_{max}}+7/2)\hbar \omega }$. A comparison between a more conservative
definition $\lambda =\sqrt{M\hbar \omega }$ and $\lambda =\sqrt{\frac{%
M\omega }{4(N_{max}+7/2)}}$ can be found in Ref.\cite{coon}. For the purpose
of our discussion here, the exact form of the infrared cutoff does not play
a significant role. The main advantage of our approach is that the
truncation is only controlled by two scales ($\lambda ,\Lambda $), and this
allows a straight forward renormalization group analysis. Moreover, recent
studies\cite{NTvK05,wpcfail} suggest that the subleading chiral potential
cannot be included non-perturbatively, if one requires the result to be
renormalization group (RG) invariant. Therefore, in this work we adopt the
new power counting\cite{Birse,Valdper,BY11} which treats the subleading
order potential perturbatively and indicate how the perturbative treatment
of chiral potential can be applied in few- and many-body calculations. Note
that although the exact power counting in chiral EFT at NN sector is still
an open question, our method can be applied to any arrangement of power
counting.

The structure of the present work is as follows. In section II, we introduce
the inverse J-matrix method. In section III, we apply the J-matrix method to
the renormalization of the leading order chiral EFT potential. In section
IV, we apply chiral EFT potential up to NNLO and show how to extract its
perturbative contribution. Finally, we summarize our findings in section V.

\section{J-matrix method}

The J-matrix method was initially derived in atomic physics\cite{j1}, and
later in nuclear physics using harmonic oscillator basis\cite{j2,j3}. The
main idea is to expression the asymptotic scattering wavefunction in terms
of infinite series of a chosen basis. In the following we just provide the
necessary formula used in our calculation. A more detail derivation can be
found in Ref.\cite{jisp}.

The Schr\"{o}dinger equation reads:%
\begin{equation}
H\Psi =(-\frac{\nabla ^{2}}{2\mu }+V)\Psi =E\Psi .  \label{eq:2.1}
\end{equation}%
Here $\mu =M/2$ is the reduced mass of the NN system and $V$ represents the
NN potential.

After partial-wave decomposition, the radial part of wavefunction can be
expanded in HO basis: 
\begin{equation}
\Psi _{l}(r)=\frac{u_{l}(r)}{r}=\sum_{n=0}^{n_{max}}c_{nl}\phi _{nl}(r).
\label{eq:2.d}
\end{equation}%
Here $l$ denotes the angular momentum quantum number, $c_{nl}$ are constants
and we have truncated the model space to $n_{\max }$ shells. The HO
wavefunction $\phi _{nl}(r)$ reads%
\begin{equation}
\phi _{nl}(r)=(-1)^{n}[2\pi \Gamma
(l+3/2)]^{-1/2}b^{-3/2}[L_{n}^{(l+1/2)}(0)]^{-1/2}(\frac{r}{b})^{l}\exp [-%
\frac{r^{2}}{2b^{2}}]L_{n}^{(l+1/2)}(\frac{r^{2}}{b^{2}}),  \label{eq:3.8}
\end{equation}%
with $b=\frac{1}{\sqrt{\mu \omega }},$ $L_{n}^{(\alpha )}$ the generalized
Laguerre polynomial. $\phi _{nl}(r)$ satisfies%
\begin{eqnarray*}
\int_{0}^{\infty }\phi _{nl}^{2}(r)4\pi r^{2}dr &=&1, \\
(2n+l+\frac{3}{2})\omega &=&\int_{0}^{\infty }\phi _{nl}(r)[\frac{1}{2\mu }(-%
\frac{1}{r}\frac{d^{2}}{dr^{2}}r+\frac{l(l+1)}{r^{2}})+\frac{1}{2}\mu
r^{2}\omega ^{2}]\phi _{nl}(r)4\pi r^{2}dr.
\end{eqnarray*}%
The maximum accessible energy in the model space is characterized by the
ultraviolet cutoff $\Lambda =\sqrt{M(2n_{_{max}}+l+3/2)\hbar \omega }$.

The kinetic energy under HO basis reads 
\begin{eqnarray}
T_{n,n-1}^{l} &=&-\frac{1}{2}\sqrt{n(n+l+1/2)},  \label{eq:3.6} \\
T_{n,n}^{l} &=&\frac{1}{2}(2n+l+3/2), \\
T_{n,n+1}^{l} &=&-\frac{1}{2}\sqrt{(n+1)(n+l+3/2)}, \\
T_{n,m}^{l} &=&0\text{ (for }|n-m|\geq 2\text{)}.
\end{eqnarray}

For the potential, we adopt the momentum space form, and one has%
\begin{equation}
V_{ll^{\prime }}(r,r^{\prime })=\frac{2}{\pi }\int_{0}^{\infty
}k^{2}dk\int_{0}^{\infty }p^{2}dpj_{l}(kr)V_{ll^{\prime }}(k,p,\Lambda
)j_{l^{\prime }}(pr^{\prime }),
\end{equation}%
where $l(l^{\prime })$ represents the angular momentum quantum numbers. The
momentum space potential is given by%
\begin{equation}
V_{ll^{\prime }}(k,p,\Lambda )=\left[ V_{ll^{\prime
}}^{LR}(k,p)+V_{ll^{\prime }}^{SR}(k,p)\right] R(k,p,\Lambda ),  \label{lop}
\end{equation}%
where the superscript $LR$ ($SR$) denotes the long- (short-) range part of
the potential. $R(k,p,\Lambda )$ is a regulator, and in this work we adopt%
\begin{equation}
R(k,p,\Lambda )=\exp (-\frac{p^{4}+k^{4}}{\Lambda ^{4}}).
\end{equation}%
Once the coordinate space representation $V_{ll^{\prime }}(r,r^{\prime })$
is obtained, the matrix element of Hamiltonian $H_{ll^{\prime }}$ reads:%
\begin{equation}
\langle H_{ll^{\prime }}\rangle _{nm}=T_{n,m}^{ll^{\prime
}}+\int_{0}^{\infty }4\pi r^{2}dr\int_{0}^{\infty }4\pi r^{\prime
2}dr^{\prime }\phi _{nl}(r)V_{ll^{\prime }}(r,r^{\prime })\phi _{ml^{\prime
}}(r^{\prime }),  \label{2.11}
\end{equation}%
with%
\begin{equation}
T_{n,m}^{l\neq l^{\prime }}=0.
\end{equation}

Moreover, due to the special property of the HO potential, one can further
simplify the above equation into: 
\begin{eqnarray}
\langle H_{ll^{\prime }}\rangle _{nm} &=&T_{n,m}^{ll^{\prime }}+\frac{2}{\pi 
}\int_{0}^{\Lambda _{c}}k^{2}dk\int_{0}^{\Lambda _{c}}p^{2}dp\phi
_{nl}(k)V_{ll^{\prime }}(k,p,\Lambda )\phi _{ml^{\prime }}(p)  \label{v} \\
&=&T_{n,m}^{ll^{\prime }}+V_{n,m}^{ll^{\prime }},
\end{eqnarray}%
where $\phi _{nl}(k)$ has the same form as $\phi _{nl}(r)$, with $r$
replaced by $k$ and $b=\sqrt{\mu \omega }$. The energy spectrum, $E_{n},$
can be obtained by diagonalizing $H_{ll^{\prime }}$ (for coupled-channels,
all possible $ll^{\prime }$ need to be included). Note that although the
regulator $R(k,p,\Lambda )$ alone is sufficient for the integral to
converge, we also impose an additional sharp cutoff $\Lambda _{c}$ (which is
set to $\Lambda _{c}=\Lambda +200$ MeV throughout this work) in Eq.~(\ref{v}%
) just to reduce the numerical task. The ultraviolet property of the
potential is majorly determined by the intrinsic cutoff $\Lambda $.

The key of connecting the NN scattering phase shifts to $E_{n}$ is to
evaluate the potential $<V_{ll^{\prime }}>_{nm}$ up to $n=m=n_{max}$, but
keep the size of kinetic part to infinity. Then formula connecting the
asymptotic wavefunction to scattering phase shift can be shown to have the
following form:%
\begin{equation}
\tan \delta (E)=-\frac{%
S_{n_{max}l}(E)-G_{n_{max}n_{max}}T_{n_{max},n_{max}+1}^{l}S_{n_{max}+1,l}(E)%
}{%
C_{n_{max}l}(E)-G_{n_{max}n_{max}}T_{n_{max},n_{max}+1}^{l}C_{n_{max}+1,l}(E)%
},  \label{un1}
\end{equation}%
with%
\begin{eqnarray}
S_{nl}(E) &=&\sqrt{\frac{\pi bn!}{\Gamma (n+l+3/2)}}(\frac{2E}{\omega })^{%
\frac{l+1}{2}}\exp [-\frac{E}{\omega }]L_{n}^{l+1/2}(\frac{2E}{\omega }),
\label{un2} \\
C_{nl}(E) &=&\sqrt{\frac{\pi bn!}{\Gamma (n+l+3/2)}}\frac{(\frac{2E}{\omega }%
)^{-l/2}}{\Gamma (-l+1/2)}\exp [-\frac{E}{\omega }]\Phi (-n-l-1/2,-l+1/2;%
\frac{2E}{\omega }),  \label{un3} \\
G_{nn^{\prime }} &=&-\sum_{\lambda ^{\prime }=0}^{n_{max}}\frac{\langle
n|\lambda ^{\prime }\rangle \langle \lambda ^{\prime }|n^{\prime }\rangle }{%
E_{\lambda ^{\prime }}-E}.  \label{un4}
\end{eqnarray}%
Here $\Phi (a,b;z)$ is the confluent hypergeometric function of 1$^{st}$
kind, $\langle n|\lambda ^{\prime }\rangle $ and $E_{\lambda ^{\prime }}$
are eigenvector and eigenvalues of the Hamiltonian $\langle H_{ll^{\prime
}}\rangle _{nm}$ truncated up to $n=m=n_{max.}$

\bigskip For cases where two partial-waves with the same total angular
momentum quantum number $J$ couple together, the number of basis increase to 
$N_{\max }=n_{\max }^{s}+n_{\max }^{d}+1$\footnote{%
Both s and d channel\ runs from 0 to $n_{\max }^{s,d}$, so $N_{\max
}=n_{\max }^{s}+n_{\max }^{d}+1.$}. Note that here we have labelled the two
channels as $s$ ($l=J-1$) and $d$ ($l=J+1$) here. Define%
\begin{equation}
\varsigma _{\Gamma \Gamma ^{\prime }}=-\sum_{\lambda ^{\prime }=0}^{N_{max}}%
\frac{\langle n_{\max }^{\Gamma }|\lambda ^{\prime }\rangle \langle \lambda
^{\prime }|n_{\max }^{\Gamma ^{\prime }}\rangle }{E_{\lambda ^{\prime }}-E},
\label{co1}
\end{equation}%
where $\langle n_{\max }^{\Gamma ^{(\prime )}}|\lambda ^{\prime }\rangle $
is the $\Gamma ^{(\prime )}-$wave component of the eigenvector. Then the
relation analog to Eq.~(\ref{un1})-(\ref{un4}) can be obtained by solving
the following equations,%
\begin{eqnarray}
\varsigma _{ss} &=&\frac{\Delta _{ss}(E)}{T_{N_{s},N_{s}+1}^{s}\Delta (E)},
\label{co2} \\
\varsigma _{dd} &=&\frac{\Delta _{dd}(E)}{T_{N_{d},N_{d}+1}^{d}\Delta (E)},
\label{co3} \\
\varsigma _{sd} &=&\varsigma _{ds}=\frac{K_{sd}}{%
2T_{N_{s},N_{s}+1}^{s}T_{N_{d},N_{d}+1}^{d}\Delta (E)}.  \label{co4}
\end{eqnarray}%
\begin{eqnarray}
\Delta _{ss}(E) &=&\left[ S_{n_{\max }^{s},s}(E)+K_{ss}(E)C_{n_{\max
}^{s},s}(E)\right] \left[ S_{n_{\max }^{d}+1,d}(E)+K_{dd}(E)C_{n_{\max
}^{d}+1,d}(E)\right]   \notag \\
&&-K_{sd}^{2}(E)C_{n_{\max }^{s},s}(E)C_{n_{\max }^{d}+1,d}(E),  \label{co5}
\\
\Delta _{dd}(E) &=&\left[ S_{n_{\max }^{s}+1,s}(E)+K_{ss}(E)C_{n_{\max
}^{s}+1,s}(E)\right] \left[ S_{n_{\max }^{d},d}(E)+K_{dd}(E)C_{n_{\max
}^{d},d}(E)\right]   \notag \\
&&-K_{sd}^{2}(E)C_{n_{\max }^{s}+1,s}(E)C_{n_{\max }^{d},d}(E),  \label{co6}
\\
\Delta (E) &=&\left[ S_{n_{\max }^{s}+1,s}(E)+K_{ss}(E)C_{n_{\max
}^{s}+1,s}(E)\right] \left[ S_{n_{\max }^{d}+1,d}(E)+K_{dd}(E)C_{n_{\max
}^{d}+1,d}(E)\right]   \notag \\
&&-K_{sd}^{2}(E)C_{n_{\max }^{s}+1,s}(E)C_{n_{\max }^{d}+1,d}(E).
\label{co7}
\end{eqnarray}%
\begin{eqnarray}
K_{ss}(E) &=&\frac{\tan \delta _{s}+\tan ^{2}\varepsilon \tan \delta _{d}}{%
1-\tan ^{2}\varepsilon \tan \delta _{s}\tan \delta _{d}},  \label{co8} \\
K_{dd}(E) &=&\frac{\tan \delta _{d}+\tan ^{2}\varepsilon \tan \delta _{s}}{%
1-\tan ^{2}\varepsilon \tan \delta _{s}\tan \delta _{d}},  \label{co9} \\
K_{sd}(E) &=&K_{ds}(E)=\frac{\tan \varepsilon }{\cos \delta _{s}\cos \delta
_{d}(1-\tan ^{2}\varepsilon \tan \delta _{s}\tan \delta _{d})}.  \label{co10}
\end{eqnarray}

One first obtain $\varsigma _{ss,sd,dd}$ from the eigenvalues and
eigenvectors using Eq.~(\ref{co1}), then solve for $K_{ss,sd,dd}(E)$ in Eq.~(%
\ref{co2})-(\ref{co7}). Finally, the phase shifts ($\delta _{s},\delta _{d}$%
) and the mixing angle $\varepsilon $ can be solved from Eq.~(\ref{co8})-(%
\ref{co10}).

In principle, once we have the eigenvalues and eigenvectors, the above
approach allows us to obtain the phase shifts at any energy. In our
approach, we adjust the LECs in our chiral potential and perform best fit of
the resulting phase shifts to the Nijmegen analysis\cite{St93,St94}.

\bigskip

\section{Leading order results}

\begin{table}[bt]
\begin{tabular}{|c|c|}
\hline
$\mathcal{O}(1)$ & OPE,\; $C_\cs{1}{0}$,\; $%
\begin{pmatrix}
C_\cs{3}{1} & 0 \\ 
0 & 0%
\end{pmatrix}
$,\; $C_\cp{3}{0}p^{\prime}p$,\; $%
\begin{pmatrix}
C_\cp{3}{2} p^{\prime}p & 0 \\ 
0 & 0%
\end{pmatrix}%
$ \\ \hline
$\mathcal{O}(Q)$ & $D_\cs{1}{0}({p^{\prime}}^2 + p^2)$ \\ \hline
$\mathcal{O}(Q^2)$ & TPE0,\; $E_\cs{1}{0}{p^{\prime}}^2 p^2$,\; $%
\begin{pmatrix}
D_\cs{3}{1}({p^{\prime}}^2 + p^2) & E_\text{SD}\,p^2 \\ 
E_\text{SD}\,{p^{\prime}}^2 & 0%
\end{pmatrix}
$,\; \\ 
& $D_\cp{3}{0}\,p^{\prime}p({p^{\prime}}^2 + p^2)$,\; $p^{\prime}p 
\begin{pmatrix}
D_\cp{3}{2}({p^{\prime}}^2 + p^2) & E_\text{PF}\,p^2 \\ 
E_\text{PF}\,{p^{\prime}}^2 & 0%
\end{pmatrix}
$, \\ 
& $C_\cp{1}{1} p^{\prime}p$,\; $C_\cp{3}{1}p^{\prime}p$ \\ \hline
$\mathcal{O}(Q^3)$ & TPE1,\; $F_\cs{1}{0}{p^{\prime}}^2p^2({p^{\prime}}^2 +
p^2)$ \\ \hline
\end{tabular}%
\caption{Power counting for pion exchanges, $S$ and $P$-wave counterterms up
to $\mathcal{O}(Q^3)$. $p$ ($p^{\prime}$) is the magnitude of the
center-of-mass incoming (outgoing) momentum. The two-by-two matrices are for
the coupled-channels.}
\label{PC}
\end{table}

\begin{figure}[h]
\begin{center}
\includegraphics[width=8cm]{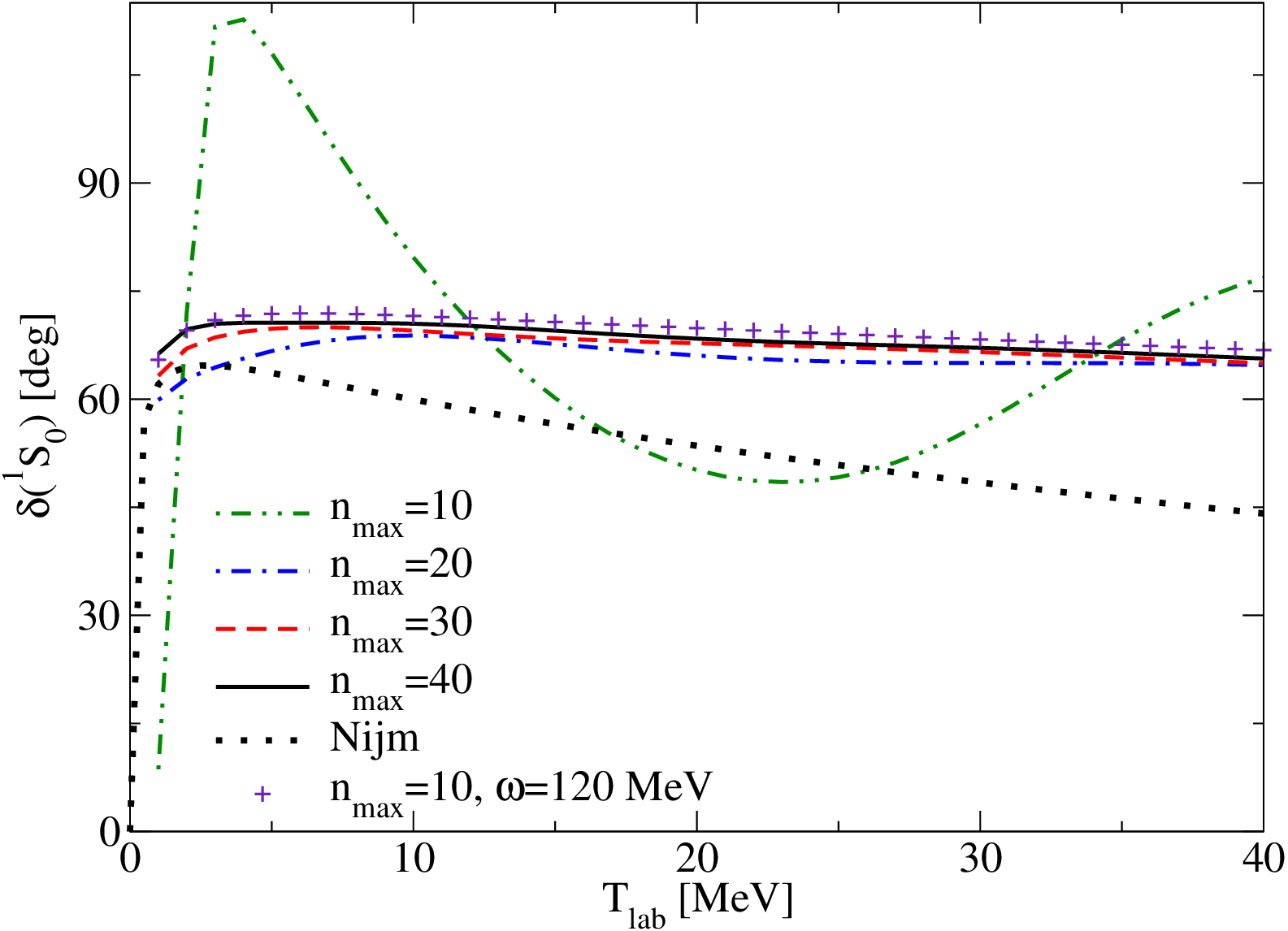}
\end{center}
\caption{{}$^{1}S_{0}$ LO phase shift as a function of laboratory energy $%
T_{lab}=0-50$ MeV. Here the black-dot represent the Nijmegen phase shift,
and each colored-line represents the phase shift with various $n_{max}$,
where $\protect\omega $ is fixed to 20 MeV except for the one with the
\textquotedblleft +" sign. Here the result is renormalized to give $%
a_{0}=-23.7$ fm, and the potential has intrinsic cutoffs $\Lambda _{c}=800$
MeV, $\Lambda =600$ MeV.}
\label{1s0plot1}
\end{figure}

The leading order potential entered in Eq.~(\ref{lop}) is the
one-pion-exchange potential (OPE). In this work we consider partial-waves $%
^{1}S_{0,}$ $^{3}S_{1}-^{3}D_{1},$ $^{1}P_{1},$ $^{3}P_{0},$ $^{3}P_{1},$ $%
^{3}P_{2}-^{3}F_{2}$. The associated contact terms are listed in Table \ref%
{PC} in terms of $O(Q^{n})$: the order where the final amplitude is summed
up to. Note that for singular attractive P-waves ($^{3}P_{0},$ $%
^{3}P_{2}-^{3}F_{2}$), the contact terms are promoted to appear one order
earlier with respect to the Weinberg power counting\cite{We90,We91}. The
contact terms, when presented, are renormalized to produce best fit to the
Nijmegen phase shifts at laboratory energy $T_{lab}\leq 10$ MeV. The two
exceptions are the $^{1}S_{0}$ and $^{3}S_{1}-^{3}D_{1}$ channel, where we
renormalize to their scattering length $a_{0}$. Once renormalization is
completed, we examine how well the NN scattering phase shifts could be
reproduced in the truncated HO space. In Fig. \ref{1s0plot1} we plot the $%
^{1}S_{0}$ phase shift obtained with $n_{max}=10-40$ and $\omega =20$ MeV.
The two cutoffs in Eq.~(\ref{v}) are set to $\Lambda _{c}=800$ MeV, $\Lambda
=600$ MeV. As one can see, with the increase of $n_{max}$---which implies
the increase of ultraviolet cutoff the decrease of the infrared cutoff $%
\lambda $---all of the LO phase shifts converge to those obtained by solving
Lippmann-Schwinger equation in the continuum.

One feature of J-matrix method can be seen in the $n_{max}=10$ curve in Fig.%
\ref{1s0plot1} is the oscillatory behavior in phase shift $\delta $. Note
that here $\sqrt{M(N_{_{max}}+3/2)\hbar \omega }=628$ MeV$>600$ MeV. Thus,
the model space's ultraviolet cutoff is already larger than the ultraviolet
cutoff in the potential. However, the $n_{max}=10$ curve shows that the
matrix element is still not saturated by enough number of basis to reproduce
the continuum properties at all $T_{lab}$. This is also observed in Ref.\cite%
{oak}. In general, we found that $\sqrt{M(N_{_{max}}+3/2)\hbar \omega }%
>R\Lambda $ is required to eliminate the oscillatory behavior, where $R$ is
a constant greater than 1. The exact value of $R$ depends on the strength
and form of interaction, we found that for OPE, $R\sim 2$.\footnote{%
We observed that the factor $R$ increases for coupled-channels and more
singular potentials.} We note that this feature is not linked to the
infrared cutoff, as one can increase $\omega $ to 120 MeV and use the same $%
n_{max}$---which increase ultraviolet and infrared cutoff at the same
time---to eliminate the oscillatory phase shift. This is shown by the curve
with plus sign in Fig.\ref{1s0plot1}.

\begin{figure}[h]
\begin{center}
\includegraphics[width=8cm]{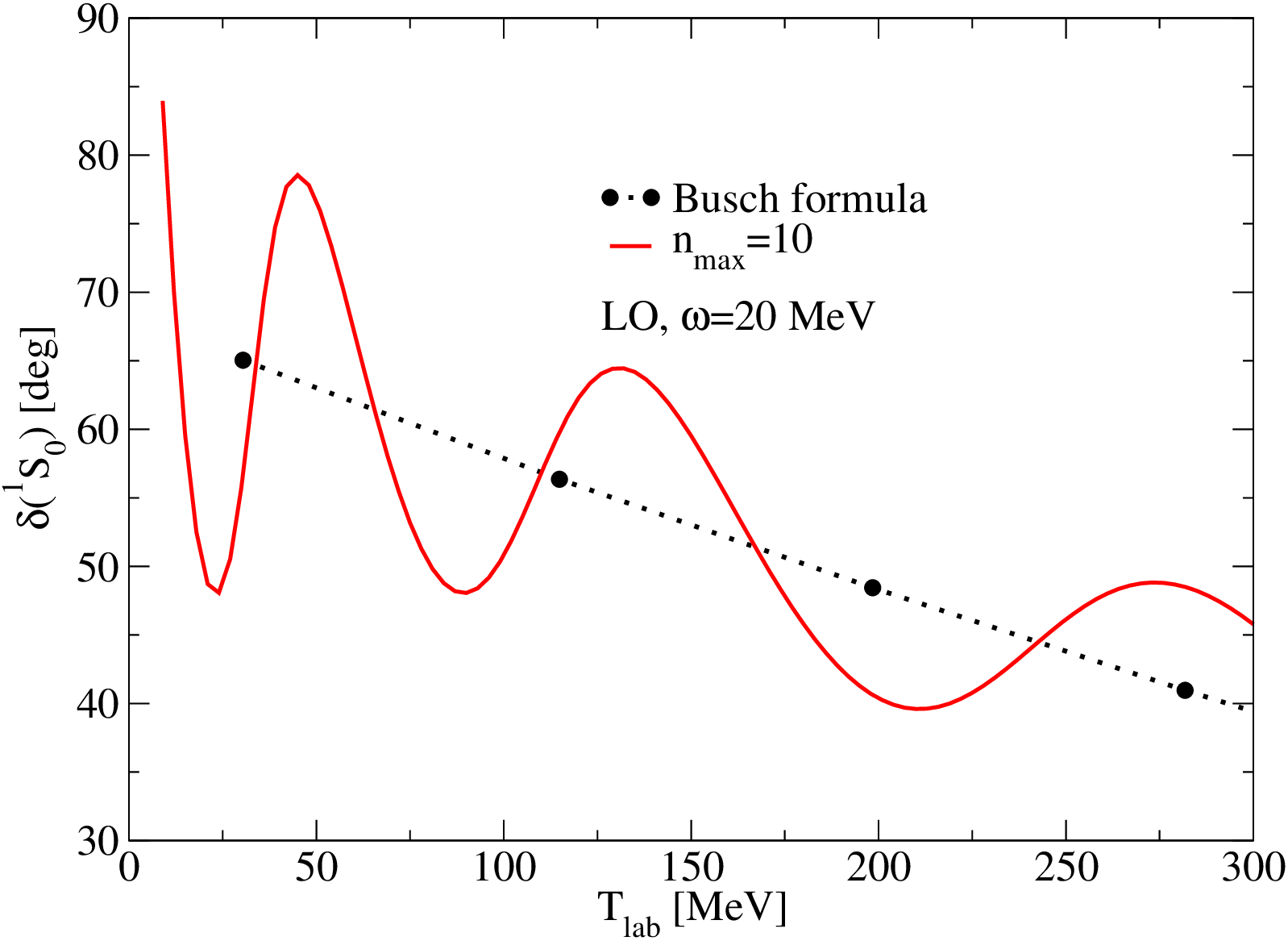}
\end{center}
\caption{{}$^{1}S_{0}$ LO phase shift as a function of laboratory energy $%
T_{lab}=0-300$ MeV. Here the black circles represent phase shift obtained by
Busch formula, and the red line represents the phase shift obtained by
J-matrix method. Here $n_{max}=10$, $\protect\omega =20$ MeV, and the
potential has intrinsic cutoffs $\Lambda _{c}=800$ MeV, $\Lambda =600$ MeV.}
\label{plot1s0_bvsj_600800}
\end{figure}

\begin{figure}[h]
\begin{center}
\includegraphics[width=8cm]{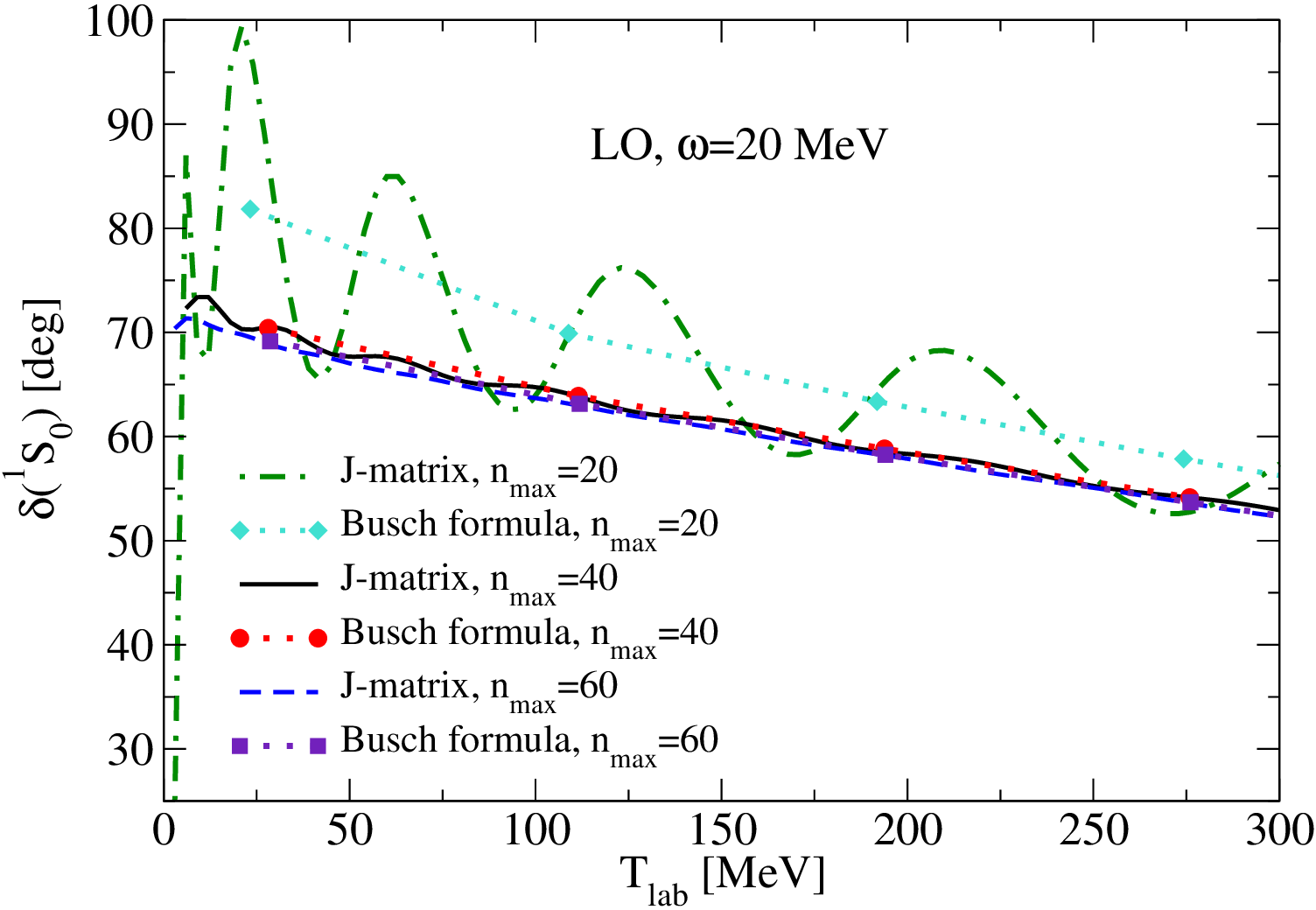}
\end{center}
\caption{{}$^{1}S_{0}$ LO phase shift as a function of laboratory energy $%
T_{lab}=0-300$ MeV. Here phase shifts obtained by Busch formula
(colored-symbol) are compared to those obtained by J-matrix method
(colored-line). Here $n_{max}=20,40,60$, and $\protect\omega =20$ MeV. The
potential has intrinsic cutoffs $\Lambda _{c}=1000$ MeV, $\Lambda =800$ MeV.}
\label{plot1s0_bvsj_w20_8001k}
\end{figure}

In order to have a further look of the problem, we insert a physical HO-trap
($\frac{1}{2}\mu r^{2}\omega ^{2}$) into the Hamiltonian. Eq.~(\ref{2.11})
then becomes 
\begin{eqnarray}
\langle H_{ll^{\prime }}\rangle _{nm} &=&T_{n,m}^{ll^{\prime }}+\frac{1}{2}%
\mu r^{2}\omega ^{2}+\int_{0}^{\infty }4\pi r^{2}dr\int_{0}^{\infty }4\pi
r^{\prime 2}dr^{\prime }\phi _{nl}(r)V_{ll^{\prime }}(r,r^{\prime })\phi
_{ml^{\prime }}(r^{\prime }) \\
&=&\delta _{ll^{\prime }}\delta _{nm}(2n+l+\frac{3}{2})\omega
+\int_{0}^{\infty }4\pi r^{2}dr\int_{0}^{\infty }4\pi r^{\prime 2}dr^{\prime
}\phi _{nl}(r)V_{ll^{\prime }}(r,r^{\prime })\phi _{ml^{\prime }}(r^{\prime
}).
\end{eqnarray}%
Here for $V_{ll^{\prime }}$ we insert the same LO potential renormalized by
the J-matrix method. Then Busch formula can then be adopted to extract the
phase shift. The result is presented in Fig.\ref{plot1s0_bvsj_600800}. As
one can see, the phase shift obtained by J-matrix method ($\delta _{J-matrix}
$) oscillates around the one obtained by Busch formula ($\delta _{busch}$).
For every c.m. energy $E_{cm}\sim 2\omega $ ($T_{lab}=2E_{cm}$), one cycle
of oscillation is completed. To show this is not a coincident, we increase
the intrinsic cutoffs of potential to $\Lambda _{c}=1000$ MeV, $\Lambda =800$
MeV and present the same comparison in Fig.\ref{plot1s0_bvsj_w20_8001k}. Fig.%
\ref{plot1s0_bvsj_600800} and \ref{plot1s0_bvsj_w20_8001k} confirm that the
oscillatory phase shift given by J-matrix method is an artifact of using
(not enough) HO-basis to represent continuum properties. The phase shift
between the two intervals just cannot be trusted. Therefore, in J-matrix
method without sufficient combination of $N_{_{max}}$ and $\omega $, one
needs to carefully choose the energies where the renormalization is
performed. Otherwise, additional error would appear due to adopting
un-trustable $\delta _{J-matrix}(E)$ in the renormalization procedure.

\begin{figure}[h]
\begin{center}
\includegraphics[width=10cm]{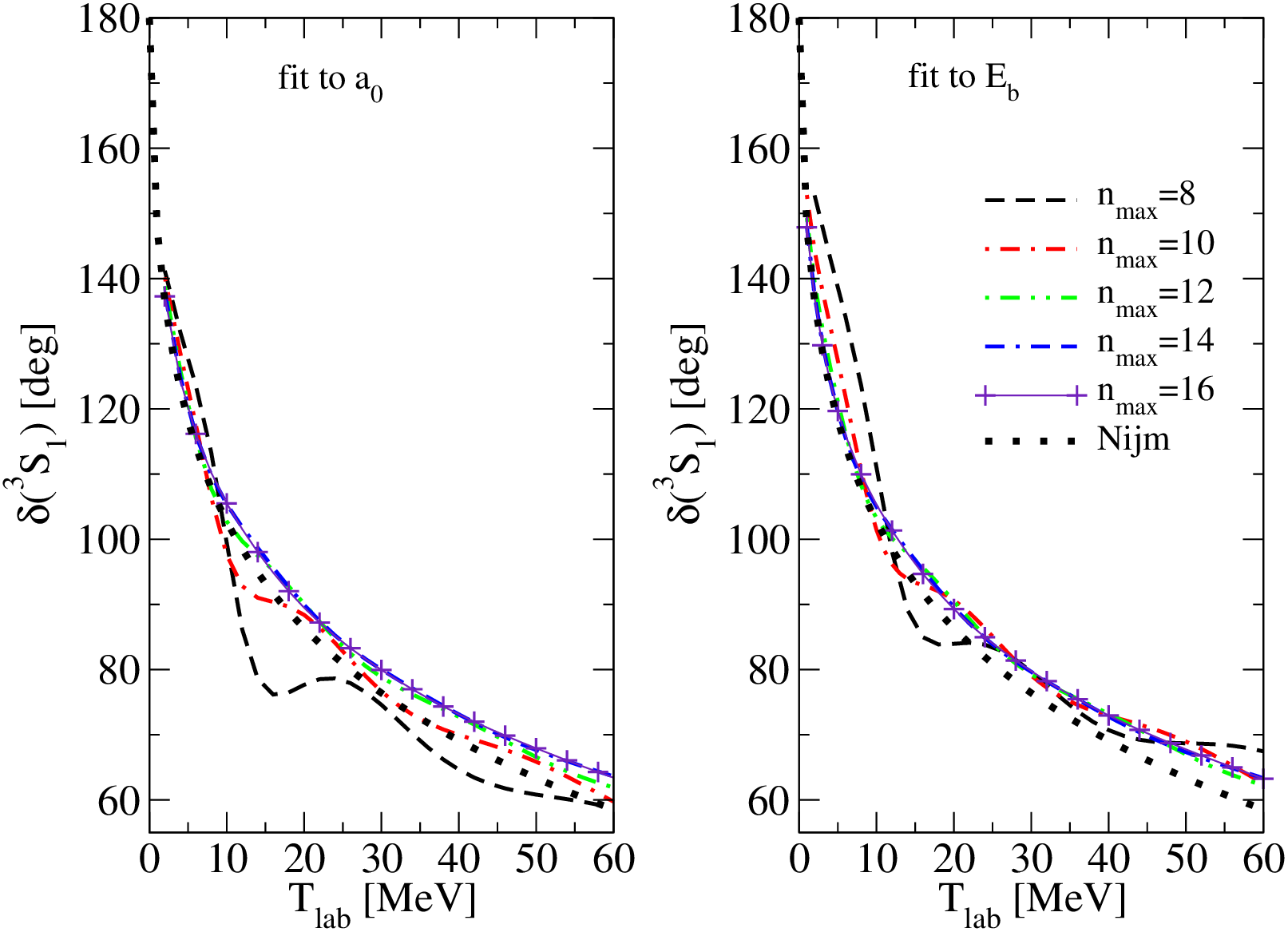}
\end{center}
\caption{{}$^{3}S_{1}$ LO phase shift as a function of laboratory energy $%
T_{lab}=0-60$ MeV. Here the black-dot represent the Nijmegen phase shift,
and each colored-line represents the phase shift with various $n_{max}$,
where $\protect\omega $ is fixed to 20 MeV. The LEC is renormalized to
reproduce $a_{0}=5.4$ fm in the left panel, and is renormalized to reproduce
the deuteron binding energy $E_{b}=-2.225$ MeV in the right panel. The
potential has intrinsic cutoffs $\Lambda _{c}=800$ MeV, $\Lambda =600$ MeV.}
\label{3s1_com}
\end{figure}

The effect of $\sqrt{M(N_{_{max}}+3/2)\hbar \omega }<R\Lambda $ appears to
be less problematic for bound-state-related properties. In Fig.\ref{3s1_com}
, we compare the $^{3}S_{1}$ phase shift with the LEC fixed by the
scattering length $a_{0}=5.4$ fm to the one fixed by and the deuteron
binding energy $E_{b}=-2.225$ MeV. From the converge pattern $n_{max}=8$ to $%
n_{max}=16$, one clearly sees that the oscillation is more centralized to
its final converged value ($n_{max}=16$ curve) in the right panel of Fig.\ref%
{3s1_com} than in the left panel. This shows that bound state indeed acts as
one of the energies where $\delta _{busch}(E)=\delta _{J-matrix}(E)$.
\bigskip 

The role of the additonal cutoff $\Lambda _{c}$ in Eq.~(\ref{v}) is just to
provide the numerical definiteness. We have verified that the phase shifts
presented above are almost unchanged (relative difference $<1\%$) by
replacing the above $\Lambda _{c}$ $(800$ MeV or $1000$ MeV) by $\Lambda
_{c}\rightarrow \infty $.

\begin{figure}[h]
\begin{center}
\includegraphics[width=12cm]{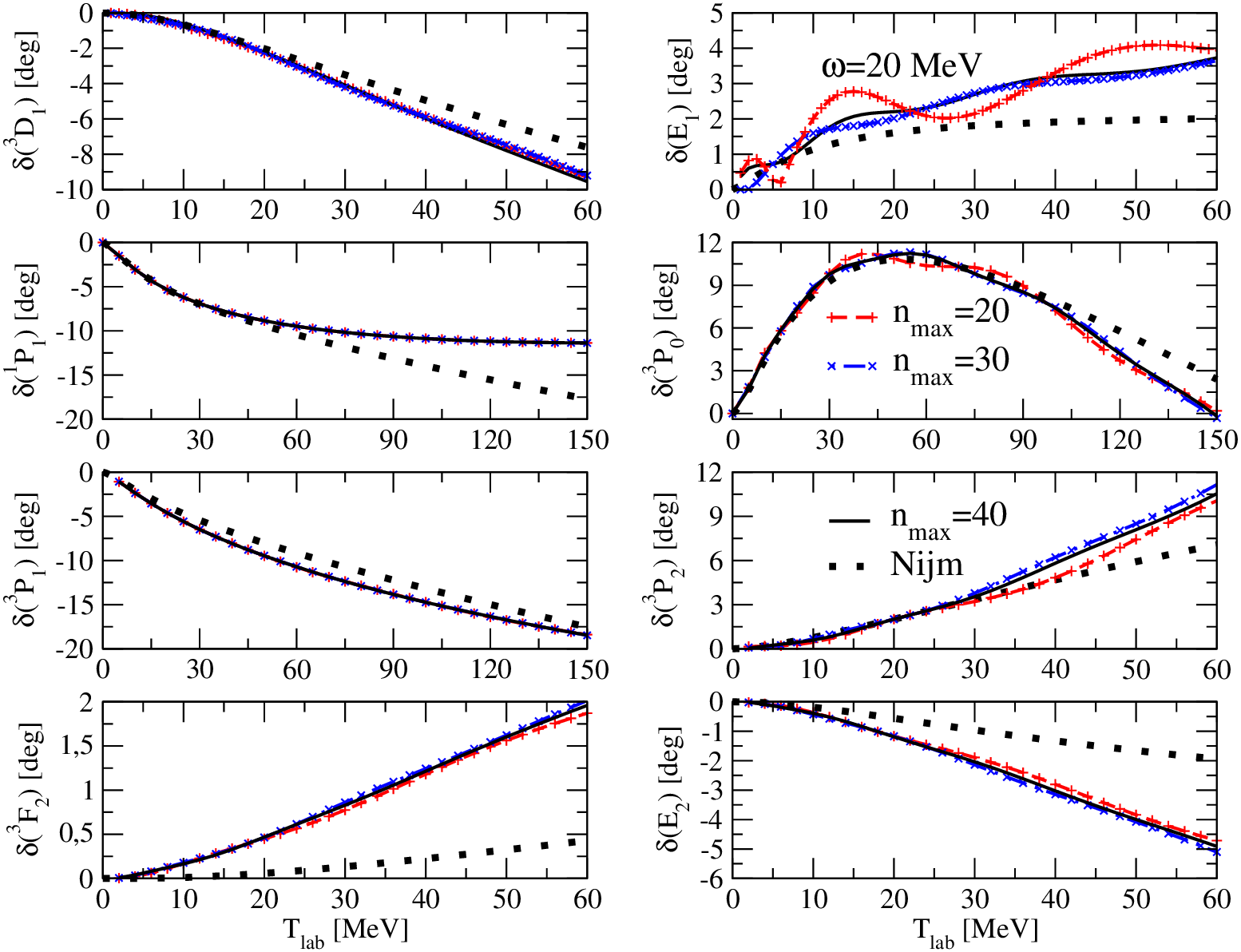}
\end{center}
\caption{Leading order phase shifts as a function of laboratory energy $%
T_{lab}$. Here the black-dot represent the Nijmegen phase shift, and each
colored-line represents the phase shift with various $n_{max}$, where $%
\protect\omega $ is fixed to 20 MeV, and the potential has intrinsic cutoffs 
$\Lambda _{c}=800$ MeV, $\Lambda =600$ MeV.}
\label{plot_lo}
\end{figure}

The $^{3}D_{1}$, $\varepsilon _{1}$ and p-waves phase shifts are shown in
Fig. \ref{plot_lo}. They present similar converge pattern as shown in
S-waves.

\begin{figure}[h]
\begin{center}
\includegraphics[width=12cm]{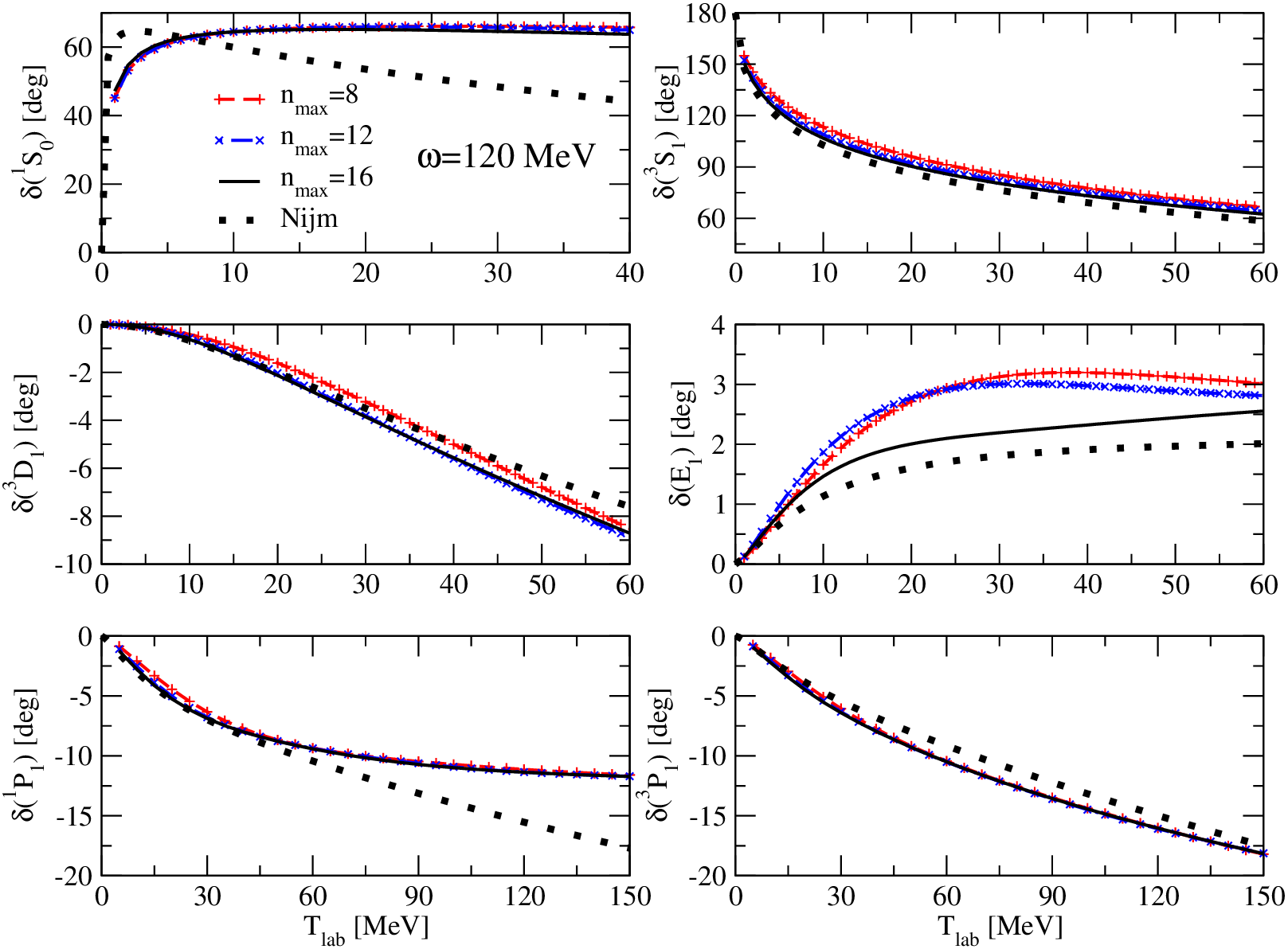}
\end{center}
\caption{Leading order phase shifts as a function of laboratory energy $%
T_{lab}$. Here the black-dot represent the Nijmegen phase shift, and each
colored-line represents the phase shift with various $n_{max}$, where $%
\protect\omega $ is fixed to 120 MeV, and the potential has intrinsic
cutoffs $\Lambda _{c}=1000$ MeV, $\Lambda =800$ MeV.}
\label{plot_lo_8001k_120}
\end{figure}

\begin{figure}[h]
\begin{center}
\includegraphics[width=12cm]{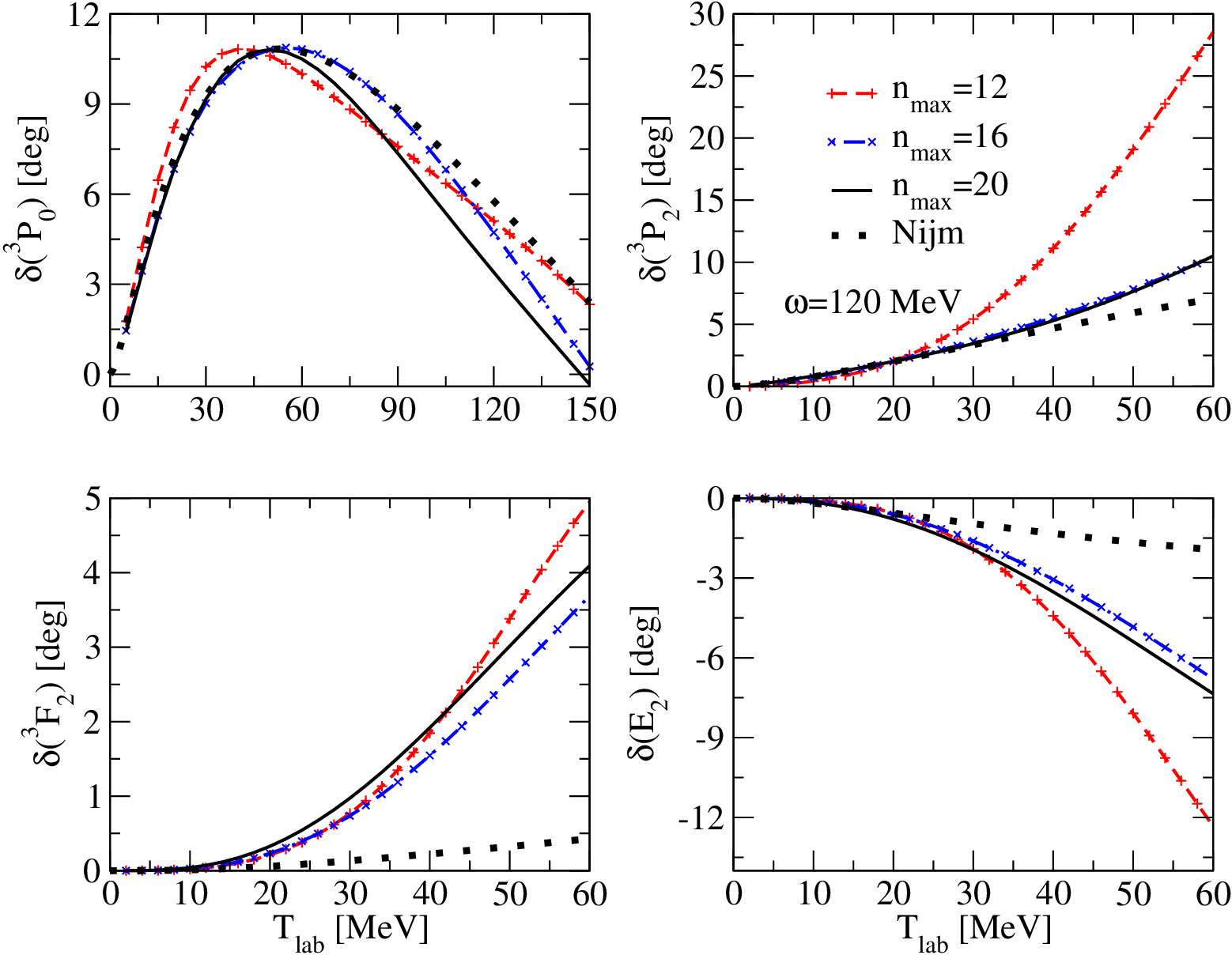}
\end{center}
\caption{Leading order phase shifts as a function of laboratory energy $%
T_{lab}$. Here the black-dot represent the Nijmegen phase shift, and each
colored-line represents the phase shift with various $N_{max}$, where $%
\protect\omega $ is fixed to 120 MeV, and the potential has intrinsic
cutoffs $\Lambda _{c}=1000$ MeV, $\Lambda =800$ MeV.}
\label{plot_lo_8001k_b}
\end{figure}

In Fig. \ref{plot_lo_8001k_120} and Fig. \ref{plot_lo_8001k_b}, we plot the
phase shifts generated by LO potential with larger intrinsic cutoffs, i.e., $%
\Lambda _{c}=1000$ MeV, $\Lambda =800$ MeV. Here we demonstrate that by
increasing $\omega =120$ MeV, the convergence can be reached with a much
lower $n_{max}$. This is due to that: (a) the ultraviolet part of the
potential is saturated with smaller $n_{max}$; (b) the oscillatory behavior
is reduced at the same time for larger $\omega $. In most of the channels, $%
n_{max}=8$ is enough to reach convergence. Meanwhile, the mixing angle $%
\varepsilon _{1}$ and those singular attractive P-waves ($^{3}P_{0}$ and $%
^{3}P_{2}-^{3}F_{2}$) require a higher $n_{max}(=16)$ to reach convergence.
Note that for $n_{max}=8$, $\sqrt{M(N_{_{max}}+3/2)\hbar \omega }\sim 1385$
MeV, which already exceeds the intrinsic ultraviolet cutoffs in the
potential. Therefore the rate of convergence in these channels ($\varepsilon
_{1}$, $^{3}P_{0}$ and $^{3}P_{2}-^{3}F_{2}$) appears to be more sensitive
to the residue-infrared-cutoff-dependence---the remaining cutoff dependence
after using contact terms to renormalize both the short- and long-range
physics.

\section{Perturbative treatment}

\subsection{Treatment for subleading potentials}

If one follows Weinberg power counting, all potentials are added up and
treated non-perturbatively. The renormalization would follow exactly the
same procedure as the LO performed in the previous section. However, in
order to achieve renormlization group (RG) invariant at arbitrary high $%
\Lambda _{a}$, it is shown that at least some of the subleading chiral
potential needed to be added perturbatively\cite{Birse,Valdper,BY11,BY}.
Here we adopt the new power counting proposed in Ref.\cite{BY11}, with
contact terms listed in Table \ref{PC}. Starting from next-to-leading order
(NLO), the potentials are treated perturbatively. The Hamiltonian we want to
solve then has the form%
\begin{equation}
H=H_{LO}+V^{(1)}+V^{(2)}+...,  \label{eq:1}
\end{equation}%
where $H_{LO}=H_{0}+V_{LO}$ is the part to be iterated to all order, and the
rest ($V^{(1)}+V^{(2)}+...$) are to be treated as perturbation. Here the
superscript denotes the order in perturbation theory where the potential
enters.

The corresponding wavefunction and energy are:%
\begin{equation}
\Psi =\Psi _{LO}+\Psi ^{(1)}+\Psi ^{(2)}+...,  \label{eq:2}
\end{equation}%
\begin{equation}
E=E_{LO}+E^{(1)}+E^{(2)}+....  \label{eq:3}
\end{equation}

In perturbation theory one has to solve:

\bigskip LO ($E_{LO}$):%
\begin{equation}
(H_{LO}-E_{LO})\Psi _{LO}=0
\end{equation}%
NLO ($E^{(1)}$):%
\begin{equation}
(H_{LO}-E_{LO})\Psi ^{(1)}=(E^{(1)}-V^{(1)})\Psi _{LO}
\end{equation}%
NNLO ($E^{(2)}$):%
\begin{equation}
(H_{LO}-E_{LO})\Psi ^{(2)}=(E^{(2)}-V^{(2)})\Psi _{LO}+(E^{(1)}-V^{(1)})\Psi
^{(1)}
\end{equation}

However, the above is difficult to deal with, especially when the
basis-state grows, as each higher order correction demands accurate
information from all eigen-states at previous order. Moreover, the above
procedure is difficult for the implement into few- and many-body calculation.

A better way to perform the perturbative calculation is to associate a small
parameter $\sigma^v $ to $V^{(v)}$, 
\begin{equation}
H(v,\sigma )=H_{LO}+\sigma V^{(1)}+\sigma ^{2}V^{(2)}+...+\sigma ^{v}V^{(v)},
\label{eq:7}
\end{equation}%
where $v$ denotes order of truncation. The perturbative solution, which is
the one we would like to extract, is 
\begin{equation}
E(v)=E_{LO}+E^{(1)}+E^{(2)}+...E^{(v)}.  \label{8}
\end{equation}

On the other hand, denote the full non-perturbative eigen-energy (obtained
by directly diagonalizing Eq.~(\ref{eq:7}) truncated at order $v$) by $\xi
(v,\sigma )$. One can express $\xi (v,\sigma )$ as 
\begin{equation}
\xi (v,\sigma )=E_{LO}+\sigma E^{(1)}+\sigma ^{2}E^{(2)}+...\sigma
^{v}E^{(v)}+O(\sigma ^{v+1}Q^{v+1}).  \label{eq:9}
\end{equation}%
Then by varying $\sigma $ and diagonalizing $H(v,\sigma ),$ one can extract $%
E^{(1,2,3,...)}$ in Eq.~(\ref{8}).

We note that this method is very general and can be directly applied to few-
and many-body calculations without modifying the existing codes.

\subsection{NLO and NNLO results}

\begin{figure}[h]
\begin{center}
\includegraphics[width=12cm]{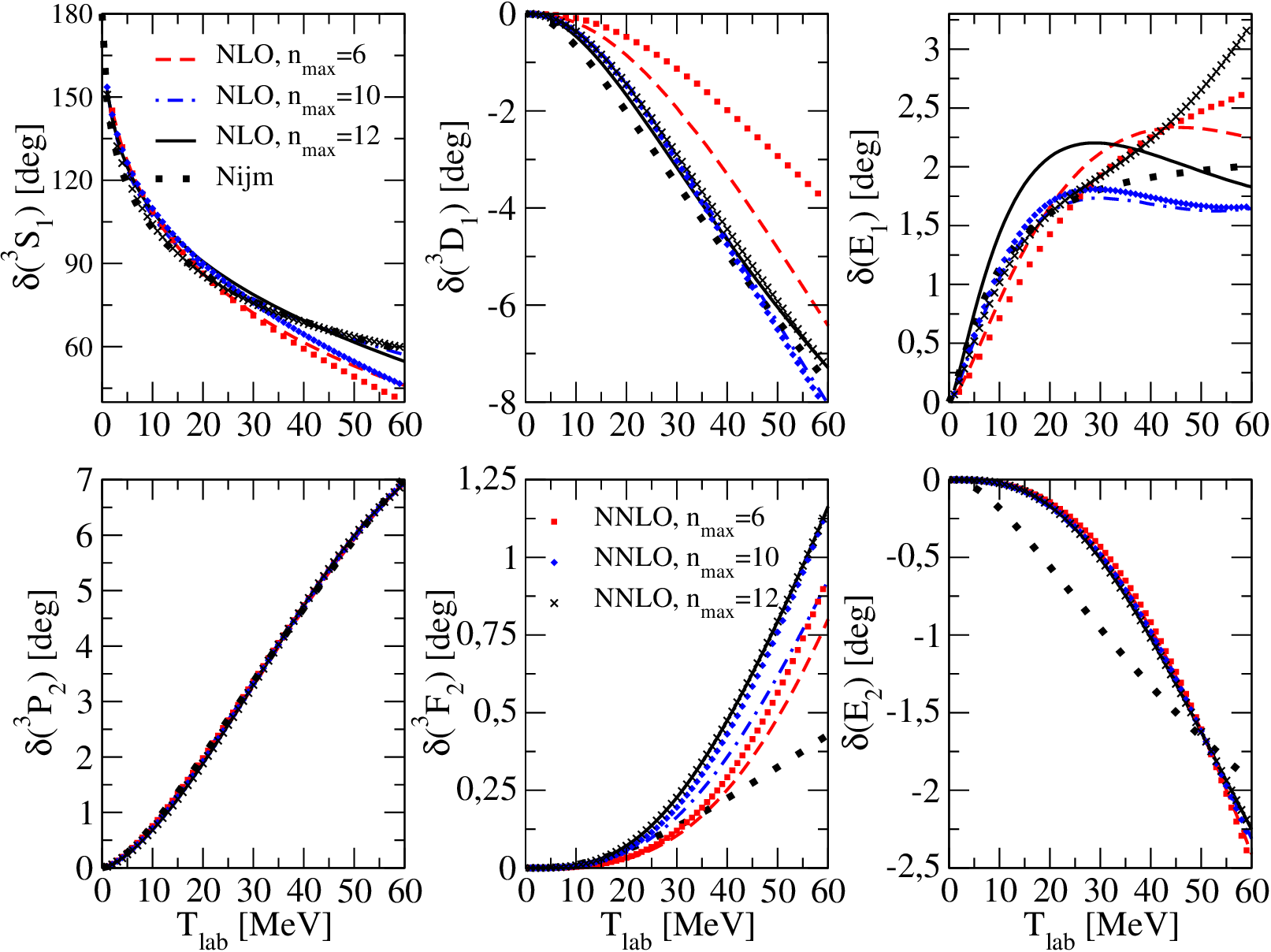}
\end{center}
\caption{NLO and NNLO coupled-channel phase shifts as a function of
laboratory energy $T_{lab}$. Here the black-dot represent the Nijmegen phase
shift, and each colored-line represents the phase shift with various $n_{max}
$, where $\protect\omega $ is fixed to 120 MeV and the intrinsic cutoffs are 
$\Lambda _{c}=800$ MeV, $\Lambda =600$ MeV.}
\label{plot_nnlo_w120}
\end{figure}

\begin{figure}[h]
\begin{center}
\includegraphics[width=12cm]{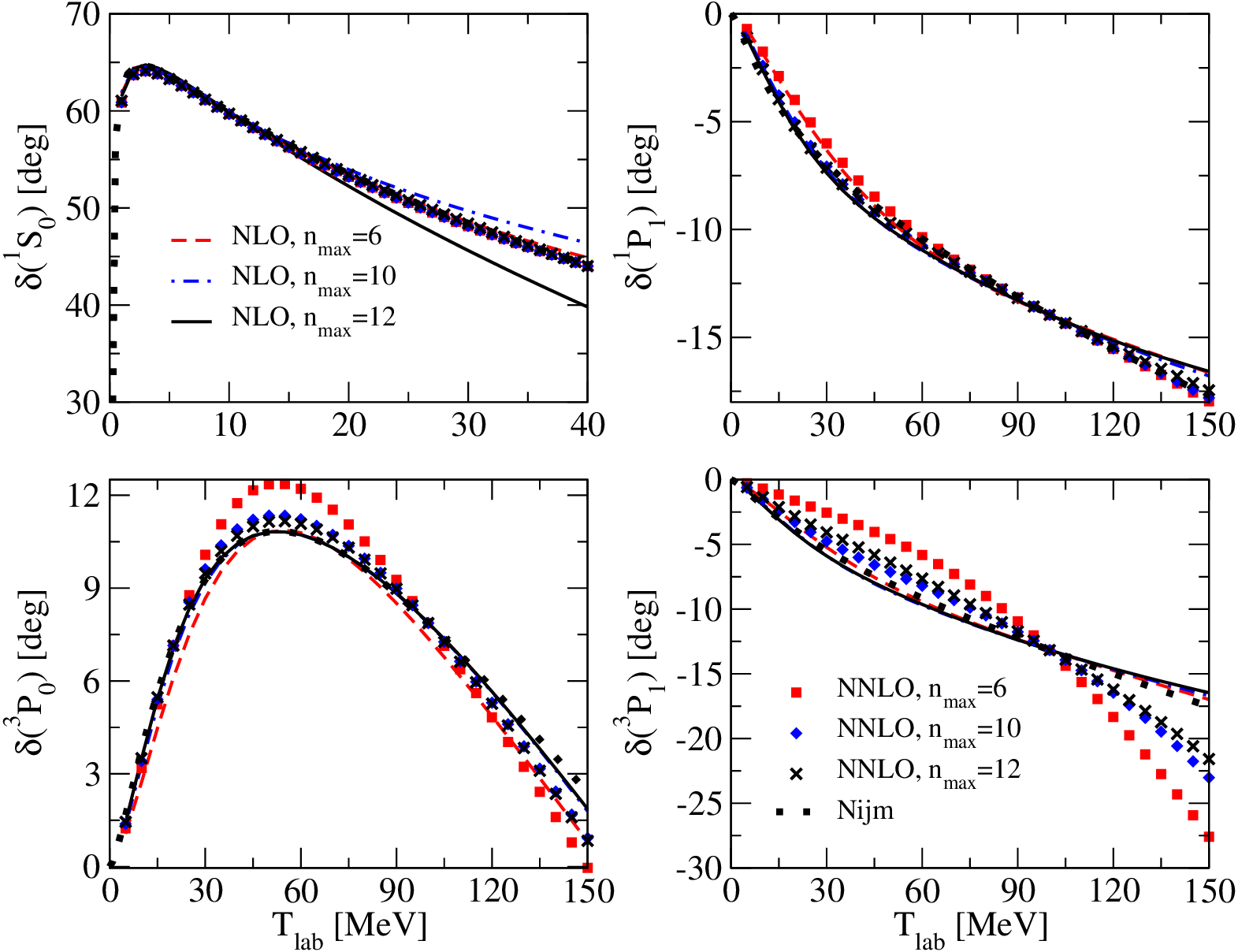}
\end{center}
\caption{NLO and NNLO uncoupled-channel phase shifts as a function of
laboratory energy $T_{lab}$. Here the black-dot represent the Nijmegen phase
shift, and each colored-line represents the phase shift with various $n_{max}
$, where $\protect\omega $ is fixed to 120 MeV and the intrinsic cutoffs are 
$\Lambda _{c}=800$ MeV, $\Lambda =600$ MeV.}
\label{plotnnlo_w120_s}
\end{figure}

\begin{figure}[h]
\begin{center}
\includegraphics[width=12cm]{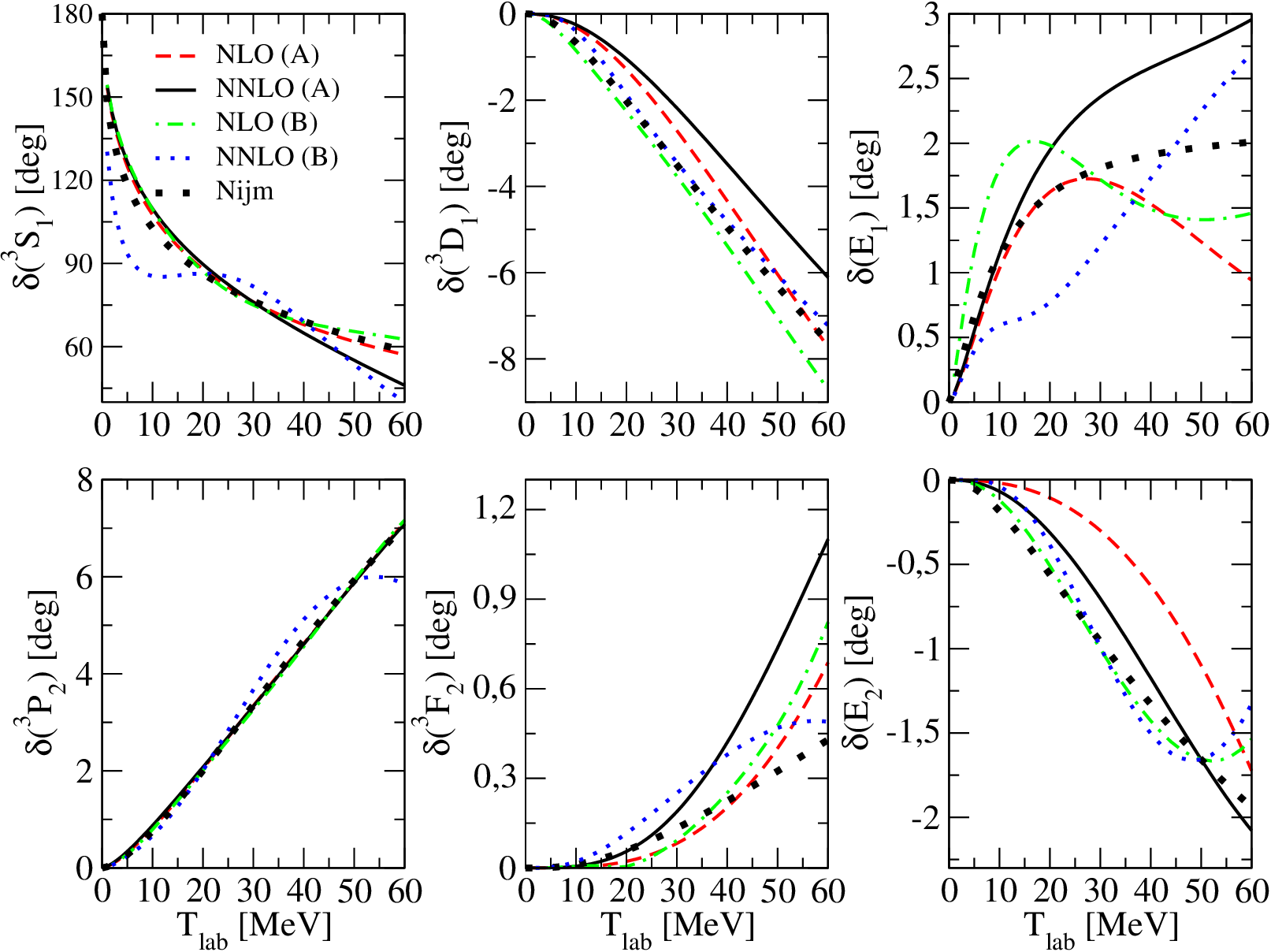}
\end{center}
\caption{NLO and NNLO coupled-channel phase shifts as a function of
laboratory energy $T_{lab}$. Here the intrinsic cutoffs are $\Lambda
_{c}=1000$ MeV, $\Lambda =800$ MeV, and black-dot represent the Nijmegen
phase shift. Each colored-line represents the phase shift at various order
and combination of ($\protect\omega ,n_{max}$). Label (A) stands for ($%
\protect\omega ,n_{max}$)=(120 [MeV],8) for $^{3}S_{1}-^{3}D_{1}$ channel
and (120 [MeV],6) for $^{3}P_{2}-^{3}F_{2}$ channel, and (B) stands for ($%
\protect\omega ,n_{max}$)=(60 [MeV],19) for $^{3}S_{1}-^{3}D_{1}$ channel
and (60 [MeV],15) for $^{3}P_{2}-^{3}F_{2}$ channel.}
\label{plot_nnlo_w60120}
\end{figure}

\begin{figure}[h]
\begin{center}
\includegraphics[width=12cm]{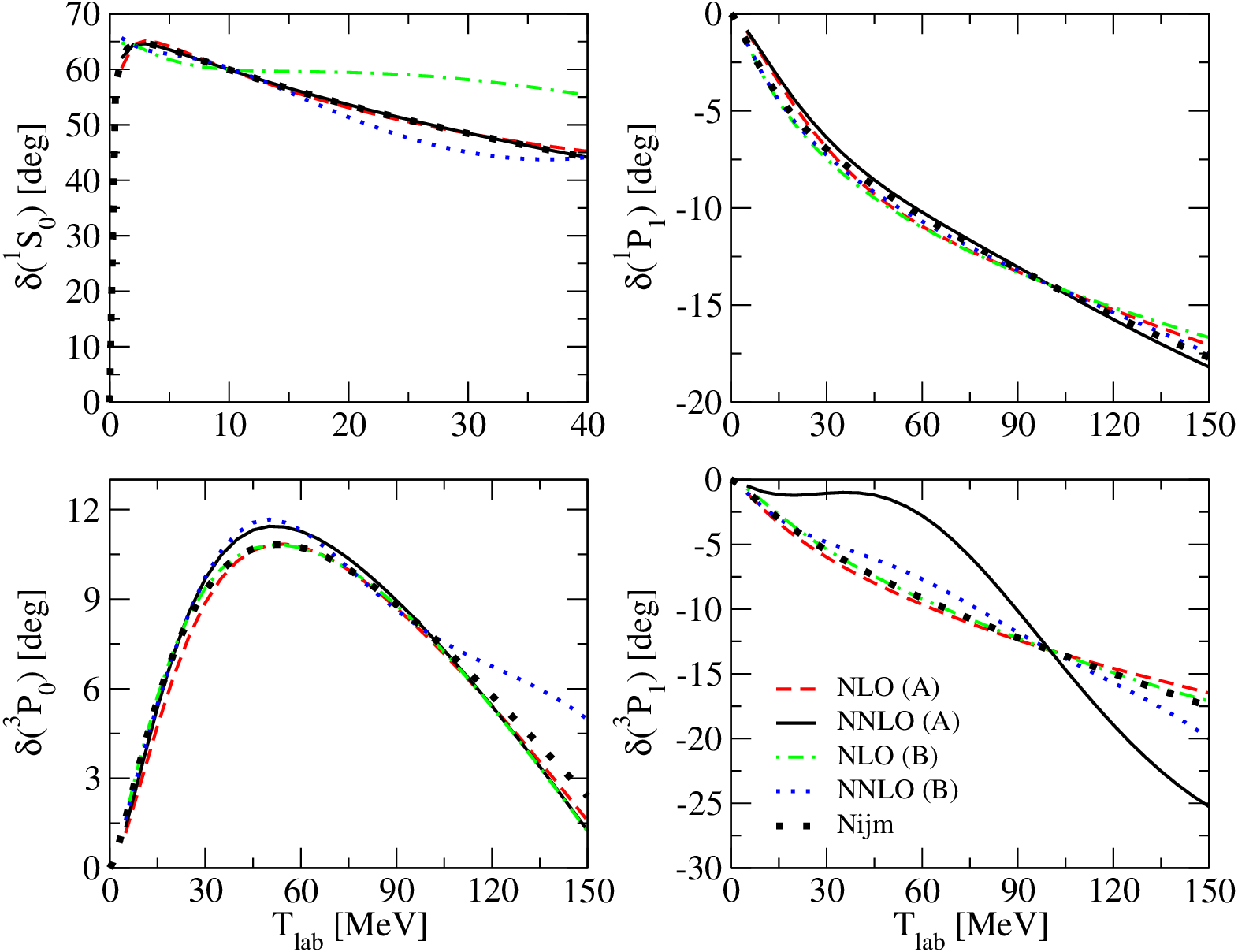}
\end{center}
\caption{NLO and NNLO uncoupled-channel phase shifts as a function of
laboratory energy $T_{lab}$. Here the intrinsic cutoffs are $\Lambda
_{c}=1000$ MeV, $\Lambda =800$ MeV, and black-dot represent the Nijmegen
phase shift. Each colored-line represents the phase shift at various order
and combination of ($\protect\omega ,n_{max}$). Label (A) stands for ($%
\protect\omega ,n_{max}$)=(120 [MeV],8), and (B) stands for ($\protect\omega %
,n_{max}$)=(60 [MeV],17).}
\label{plotnnlo_w60120_s}
\end{figure}

The NLO and NNLO phase shifts based on power counting of Ref.\cite{BY11} are
presented in Figs. \ref{plot_nnlo_w120}-\ref{plotnnlo_w60120_s}. Here Fig. %
\ref{plot_nnlo_w120} and \ref{plotnnlo_w120_s} are for potential with
intrinsic cutoffs $\Lambda _{c}=800$ MeV, $\Lambda =600$ MeV, and Fig. \ref%
{plot_nnlo_w60120} and \ref{plotnnlo_w60120_s} are for $\Lambda _{c}=1000$
MeV, $\Lambda =800$ MeV. The subleading phase shifts are obtained
perturbatively according to the method introduced in section IV A. The LECs
are renormalized to reproduce the Nijmegen phase shifts up to the maximum $%
T_{lab}$ shown in each channel. In $^{3}S_{1}-^{3}D_{1}$ channel, the
deuteron binding energy $E_{b}=-2.224$ MeV is also adopted in the fit.

Unlike the conventional Weinberg counting, where the order of chiral
potential equals to the order in the final amplitude, the same dose not
necessary hold for the new power counting. Denote the leading (subleading)
two-pion-exchange potential as TPE0 (TPE1)\footnote{%
In Weinberg counting, TPE0 equals to the NLO$(Q^{2})$ and TPE1 equals to the
NNLO$(Q^{3})$ potential.}, in most of the channels presented here, TPE0
enters at NLO and TPE1 enters at NNLO. For these cases potential up to TPE0
and TPE1 both enter as one insertion\footnote{%
This means one only extracts $E^{(1)}$ in Eq. (\ref{8}).} in the LO
wavefunction, and the resulting phase shifts are NLO and NNLO, respectively.
However, when a non-vanishing O(Q) potential appears, such as in the $%
^{1}S_{0}$ channel, NLO contains only contact terms and TPE0 enters at NNLO.
Here NNLO includes one insertion of TPE0 and two insertions\footnote{%
This is equivalent to evaluate up to $E^{(2)}$ contribution coming from $%
V^{(1)}$ in Eq. (\ref{8}).} of the O(Q) contact term.

As one can see, with the inclusion of NLO/NNLO contribution, the phase
shifts converge already at $(\omega ,n_{max})=(120,10)$ for both of the
potentials adopted here (the one with $\Lambda _{c}=800$ MeV, $\Lambda =600$
MeV and the other with $\Lambda _{c}=1000$ MeV, $\Lambda =800$ MeV). The
reproduction of Nijmegen phase shifts are comparable to those obtained in
the continuum\cite{BY11}. For lower value of $\omega $, i.e., $\omega =60$
MeV, the minimum $n_{max}$ required for the NNLO phase shifts to converge
ranges from $n_{max}=15-20$ depends on the channels. In general, the quality
of fit we obtained are comparable to those obtained by the standard Weinberg
counting, with only one exception: the NNLO $^{3}P_{1}$ channel. In this
case, the same behavior is observed in the continuum as well\cite{BY11}.
This might suggest that the $c_{1,3,4}$ adopted in TPE1 need to be
re-adjusted, or an adoption of the $\Delta (1232)-$included potential is
necessary in order to cure this behavior.

\bigskip

\section{Conclusion}

We have performed a new approach to generate the chiral EFT potential in the
truncated model space. We utilize the contact interactions presented in EFT
to absorb the effects coming from both ultraviolet and infrared cutoffs. The
connection between eigenstate of HO-basis and NN scattering phase shift are
established by the J-matrix formalism. This allows a direct evaluation of NN
scattering in HO-basis without applying a physical trap. In our procedure,
the RG analysis can be carried out in a straight forward way as the results
depends only on two scales: the infrared cutoff $\lambda $ and the
ultraviolet cutoff $\Lambda $. This paves a way to provide a truly model
independent procedure to perform ab-initio calculations. Also, the
perturbative treatment of chiral potential is carried out in the truncated
model space through a method which is directly applicable to many-body
calculation.

There are many possibilities to extend the current study. In particular, the
interaction obtained in this work will be applied to the 3-, 4- and
many-body calculations within the no-core-shell-model framework\cite{ncsm}.


\begin{acknowledgments}
We thank A. Shirokov, N. Barnea, D. Lee, G. Hupin, M. Grasso, B.R. Barrett
T. Papenbrock, R.J. Furnstahl and U. van Kolck for useful discussions and suggestions. More importantly,
the author is grateful for the valuable discussions and supports from G.
Orlandini and W. Leidemann. Part of this work was carried out in Univ. of
Trento under MIUR grant PRIN-2009TWL3MX.
\end{acknowledgments}




\appendix



\end{document}